# *Insight*-HXMT observations of the first binary neutron star merger GW170817

TiPei Li[1,2,3], ShaoLin Xiong[1], ShuangNan Zhang[1,3*], FangJun Lu[1], LiMing Song[1], XueLei Cao[1], Zhi Chang[1], Gang Chen[1], Li Chen[4], TianXiang Chen[1], Yong Chen[1], YiBao Chen[2], YuPeng Chen[1], Wei Cui[1,2], WeiWei Cui[1], JingKang Deng[2], YongWei Dong[1], YuanYuan Du[1], MinXue Fu[2], GuanHua Gao[1,3], He Gao[1,3], Min Gao[1], MingYu Ge[1], YuDong Gu[1], Ju Guan[1], ChengCheng Guo[1,3], DaWei Han[1], Wei Hu[1], Yue Huang[1], Jia Huo[1], ShuMei Jia[1], LuHua Jiang[1], WeiChun Jiang[1], Jing Jin[1], YongJie Jin[5], Bing Li[1], ChengKui Li[1], Gang Li[1], MaoShun Li[1], Wei Li[1], Xian Li[1], XiaoBo Li[1], XuFang Li[1], YanGuo Li[1], ZiJian Li[1,3], ZhengWei Li[1], XiaoHua Liang[1], JinYuan Liao[1], CongZhan Liu[1], GuoQing Liu[2], HongWei Liu[1], ShaoZhen Liu[1], XiaoJing Liu[1], Yuan Liu[1], YiNong Liu[5], Bo Lu[1], XueFeng Lu[1], Tao Luo[1], Xiang Ma[1], Bin Meng[1], Yi Nang[1,3], JianYin Nie[1], Ge Ou[1], JinLu Qu[1], Na Sai[1,3], Liang Sun[1], Yin Tan[1], Lian Tao[1], WenHui Tao[1], YouLi Tuo[1,3], GuoFeng Wang[1], HuanYu Wang[1], Juan Wang[1], WenShuai Wang[1], YuSa Wang[1], XiangYang Wen[1], BoBing Wu[1], Mei Wu[1], GuangCheng Xiao[1,3], He Xu[1], YuPeng Xu[1], LinLi Yan[1,3], JiaWei Yang[1], Sheng Yang[1], YanJi Yang[1], AiMei Zhang[1], ChunLei Zhang[1], ChengMo Zhang[1], Fan Zhang[1], HongMei Zhang[1], Juan Zhang[1], Qiang Zhang[1], Shu Zhang[1], Tong Zhang[1], Wei Zhang[1,3], WanChang Zhang[1], WenZhao Zhang[4], Yi Zhang[1], Yue Zhang[1,3], YiFei Zhang[1], YongJie Zhang[1], Zhao Zhang[2], ZiLiang Zhang[1], HaiSheng Zhao[1], JianLing Zhao[1], XiaoFan Zhao[1,3], ShiJie Zheng[1], Yue Zhu[1], YuXuan Zhu[1], and ChangLin Zou[1] (The *Insight*-HXMT team)

[1] *Key Laboratory of Particle Astrophysics, Institute of High Energy Physics, Chinese Academy of Sciences, Beijing 100049, China*
[2] *Department of Physics, Tsinghua University, Beijing 100084, China*
[3] *University of Chinese Academy of Sciences, Chinese Academy of Sciences, Beijing 100049, China*
[4] *Department of Astronomy, Beijing Normal University, Beijing 100088, China*
[5] *Department of Engineering Physics, Tsinghua University, Beijing 100084, China*

Finding the electromagnetic (EM) counterpart of binary compact star merger, especially the binary neutron star (BNS) merger, is critically important for gravitational wave (GW) astronomy, cosmology and fundamental physics. On Aug. 17, 2017, Advanced LIGO and *Fermi*/GBM independently triggered the first BNS merger, GW170817, and its high energy EM counterpart, GRB 170817A, respectively, resulting in a global observation campaign covering gamma-ray, X-ray, UV, optical, IR, radio as well as neutrinos. The High Energy X-ray telescope (HE) onboard *Insight*-HXMT (Hard X-ray Modulation Telescope) is the unique high-energy gamma-ray telescope that monitored the entire GW localization area and especially the optical counterpart (SSS17a/AT2017gfo) with very large collection area (~1000 cm$^2$) and microsecond time resolution in 0.2-5 MeV. In addition, *Insight*-HXMT quickly implemented a Target of Opportunity (ToO) observation to scan the GW localization area for potential X-ray emission from the GW source. Although *Insight*-HXMT did not detect any significant high energy (0.2-5 MeV) radiation from GW170817, its observation helped to confirm the unexpected weak and soft nature of GRB 170817A. Meanwhile, *Insight*-HXMT/HE provides one of the most stringent constraints (~10$^{-7}$ to 10$^{-6}$ erg/cm$^2$/s) for both GRB170817A and any other possible precursor or extended emissions in 0.2-5 MeV, which help us to better understand the properties of EM radiation from this BNS merger. Therefore the observation of *Insight*-HXMT constitutes an important chapter in the full context of multi-wavelength and multi-messenger observation of this historical GW event.

**GW170817, BNS merger, gravitational wave electromagnetic counterpart**

**PACS number(s):**

---

*Corresponding author (email: zhangsn@ihep.ac.cn)

## 1 Introduction

The First direct observation of gravitational wave (GW) from the binary black hole (BBH) merger GW150914 by the Advanced Laser Interferometer Gravitational-Wave Observatory (LIGO) heralded the GW astronomy [1]. However, finding the electromagnetic (EM) counterpart of GW source is critically important to independently confirm the physical event produced the GW, to study the GW event in astrophysical context, to independently measure the Hubble constant, and to investigate fundamental physics (e.g. mass of graviton, Lorentz Invariance) [2-4].

Extensive EM follow-up observations have been implemented since the very first GW event GW150914 [5]. Although a possible EM counterpart candidate for GW150914 was reported by *Fermi*/GBM [6], other studies using the same GBM data [7, 8] and other observation using INTEGRAL/SPI-ACS [9] rejected this candidate. In fact, BBH merger is not expected to produce EM emission [10, 11], except in some certain conditions [e.g. 12, 13]. Therefore, the primary targets for GW-EM joint detection are compact binary mergers involving a neutron star, either binary neutron star (BNS) or black hole-neutron star mergers, which is supported by the extensive studies of short GRBs in decades [14].

The Hard X-ray Modulation Telescope (HXMT)[1], dubbed as *Insight*-HXMT after launch on June 15, 2017, was originally proposed in the 1990s, based on the Direct Demodulation Method [15, 16]. As China's first X-ray astronomical satellite, *Insight*-HXMT carries three main payloads onboard [17]: the High Energy X-ray telescope (HE, 20-250 keV, 5100 cm$^2$), the Medium Energy X-ray telescope (ME, 5-30 keV, 952 cm$^2$), and the Low Energy X-ray telescope (LE, 1-15 keV, 384 cm$^2$), as shown in Figure 1. The original primary scientific objectives of *Insight*-HXMT are: (1) to scan the Galactic Plane for revealing new transient sources and to monitor the known variable sources; (2) to observe X-ray binaries to study the dynamics and emission mechanism in strong gravitational or magnetic fields [17].

Thanks to the innovative operation of the anticoincidence detectors of HE, the main scientific objectives of *Insight*-HXMT have been extended to monitor Gamma-Ray Bursts (GRBs) and GW EM counterparts, making *Insight*-HXMT one of the most important gamma-ray monitors in the gravitational wave astronomy era.

HE consists of 18 cylindrical NaI(Tl)/CsI(Na)[2] phoswich detectors, each with a diameter of 190 mm and thickness of 3.5 mm and 40 mm for NaI and CsI respectively. NaI/CsI phoswich detector is originally designed to measure hard X-rays in 20-250 keV incident from the nominal Field of View (FoV, 1.1°×5.2° and 5.2°×5.2°) defined by slat col-

---

[1] http://www.hxmt.org
[2] Hereafter use NaI, CsI for brevity.

---

limators placed in front of the detectors (Figure 1), where CsI is used as anticoincidence detector to reduce the background of the main detector NaI. However, downloading the individual photon data registered by CsI allows it to detect gamma-rays with energies from ~200 keV up to tens of MeV which penetrate the spacecraft and payload structure and deposit energy in the CsI crystals.

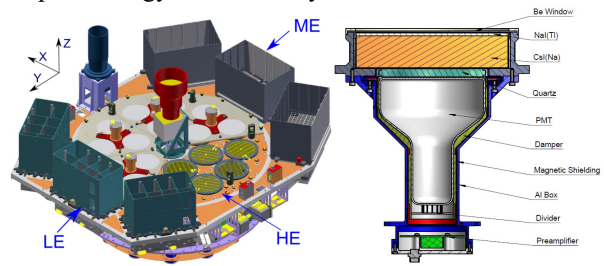

**Figure 1** (Left) *Insight*-HXMT is composed of HE, ME and LE telescopes. Coordinates used in this paper are shown on the upper left. Incident angle theta is the angle between line-of-sight to the source and the Z axis. Azimuthal angle phi equals 0 deg for the X axis, and 90 deg for the Y axis. (Right) The design of NaI/CsI detector. Slat collimators (not shown) are placed above the Be window.

HE normally operates in the regular mode, in which NaI works in the energy range of about 20-250 keV while CsI in about 80-800 keV. In order to improve the spectroscopy capability for hard GRBs (i.e. $E_{peak} \sim $ MeV), a dedicated working mode, called GRB mode, has been designed and implemented for HE. In this mode, the high voltage of the photomultiplier tube (PMT, readout device for NaI/CsI scintillator) is reduced so that the measured energy range of CsI goes up to 0.2 - 3 MeV. Note that the above energy is the deposited energy in CsI caused by incident gamma-rays with energy from ~200 keV to tens of MeV, and most of these gamma-rays only deposit part of their energies in CsI.

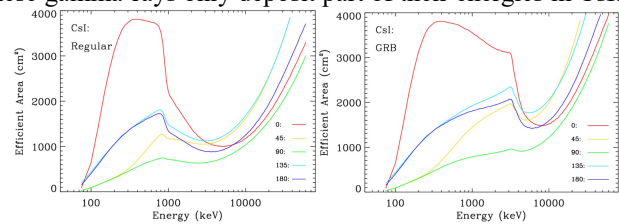

**Figure 2** Simulated total effective area of HE 18 CsI detectors for GRB detection in regular mode (Left) and GRB mode (Right). Each line represents the effective area for each theta angle averaged in azimuthal (phi angle from 0 to 360 deg).

Since the structure of the satellite is very complicated, detector response to the penetrated gamma-rays is quite difficult to compute. On-ground calibration of the response is impossible due to the constraints of the satellite. Therefore, GEANT4 [18] has been utilized to simulate the response of the CsI detector. The simulated effective area as a function of incident energy in different angles for both the regular and GRB modes are shown in Figure 2.

As shown in Figure 2, the effective area for typical incident geometry in both regular mode and GRB mode is ~1000 cm$^2$, meaning that HE is one of the largest GRB detectors in hundreds keV to tens MeV ever flown and that HE can detect GRBs in either mode. It also shows that HE is an omnidirectional detector for GRBs, thus the FoV for GRB observation is the whole sky not occulted by the Earth. Therefore, *Insight*-HXMT/HE could serve as an excellent wide-field monitor for high-energy transient sources, including GRBs and GW EM counterparts.

Since the first light on June 21, 2017, HE has detected 26 GRBs significantly and published 20 GCN Circulars[3] as of Sept. 25, 2017. Examples of the detected GRBs are shown in Figure 3. GRB 170904A (Figure 3, Left panel) is a typical bright burst, for which a very high statistics was achieved by *Insight*-HXMT, allowing us to explore the variability on the minimum timescale in MeV energy band.

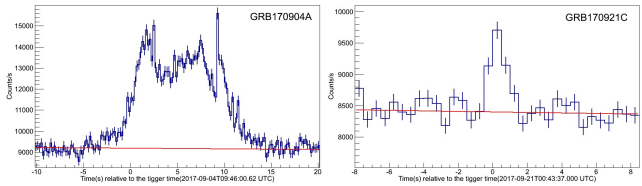

**Figure 3** Light curves of GRB 170904A (Left) and GRB 170921C (Right) detected by *Insight*-HXMT/HE.

As shown in the right panel of Figure 3, GRB 170921C is a typical short hard burst which was triggered by HE with high significance (12 σ). For comparison, *Fermi*/GBM only detected it as a sub-threshold event (total significance ~ 8 σ), while INTEGRAL/SPI-ACS did not trigger it and off-line analysis shows the detection significance is about ~4 σ.

These observation facts manifested the capability and advantage of *Insight*-HXMT/HE in detecting short hard GRBs, which are the expected high-energy EM counterparts of GW sources from binary mergers involving neutron stars (e.g. neutron star-neutron star or neutron star-black hole mergers).

## 2 GW170817 and *Insight*-HXMT Observations

On Aug. 17, 2017, at 12:41:04.446 UTC, LIGO was triggered by the first neutron-star binary coalescence GW event (GW170817, [19]). About ~2 s later, a weak short GRB (GRB 170817A, [20, 21]) triggered the *Fermi*/GBM independently at 2017-08-17T12:41:06.47 UTC (hereafter T0) [20]. The GBM localization is broadly consistent with the GW location given by the LIGO-Hanford detector [20]. Soon after, this GRB was confirmed by an offline data analysis of INTEGRAL/SPI-ACS [21]. IPN triangulation from the time delay between GBM and SPI-ACS further improved the location of GRB170817A, which is well consistent with the refined GW sky region (~30 deg$^2$, 90% CL) made by the Hanford-Livingston-Virgo detector network [19]. Coincidence in both temporal and location between GW170817 and GRB170817A strongly supports that they are physically related [22].

Alerts of those observations were distributed quickly within the LVC EM follow-up community, resulting in a historically global observation campaign covering UHECR, gamma-ray, X-ray, UV, optical, IR, radio as well as neutrinos [23]. These multi-wavelength and multi-messenger observations finally lead to the first discovery of a BNS merger with gravitational wave and EM waves spanning from gamma-ray, X-ray, ultraviolet, optical, infrared to radio band [23, and references therein].

*Insight*-HXMT is the unique gamma-ray telescope that monitored the entire GW location area, especially the optical counterpart[4] (SSS17a [24]), around the trigger time, with ~1000 cm$^2$ effective area and microsecond (μs) temporal resolution at high energies (0.2-5 MeV). Not only did *Insight*-HXMT do a prompt observation (section 2.1) to GW170817 and GRB170817A, it also quickly implemented a Target of Opportunity (ToO) observation to the GW area for potential X-ray counterpart (section 2.2), as well as a dedicated ToO observation to the Crab pulsar for calibration of the detector response (section 2.3).

Given the importance of its observation, results of *Insight*-HXMT have been included in the historical paper of GW170817 [23]. Here we report the detailed observation and data analysis of *Insight*-HXMT.

### 2.1 Prompt Observation

*Insight*-HXMT was making a pre-scheduled pointed observation to GRO J1008-57 when GW170817 happened. HE monitored the whole localization region of *Fermi*/GBM, the entire LVC sky map and the final GW source position (i.e. optical counterpart SSS17a) without any occultation by the Earth at the trigger time (T0), as shown in Figure 4. We note that the shielding or occultation effect prevented many other X/gamma-ray telescopes from observing GW170817 and GRB170817A around the trigger time [23]. Indeed, *Insight*-HXMT had been monitoring the SSS17a sky area since the first light of HE (~ T0-50 day), with interruptions only from the Earth occultation of the source and the swift-off of HE when the satellite passed through the South Atlantic Anomaly (SAA) region. During the pointed observation for GRO J1008-57, *Insight*-HXMT observed SSS17a with a constant incident geometry (theta=47.3°, phi=201.1°, definition of incident angles are shown in Figure 1 caption) from T0-1.15 day to T0+0.29 day.

As shown in Figure 4 bottom panel, HE provided a continuous coverage from T0-650 s to T0+450 s in the vicinity of trigger time where a short GRB and its related emission episodes (i.e. precursor, extended emission) produced by BNS merger are likely to occur. Within T0±5000 s, the GW source was occulted by the Earth from T0-2780 s to T0-650 s and from T0+2950 s to T0+5000 s, while HE was turned off due to SAA passage from T0+450 s to T0+1110 s.

---

[3] https://gcn.gsfc.nasa.gov/gcn3_archive.html

[4] IAU identification of AT2017gfo. Also named as DLT17ck.

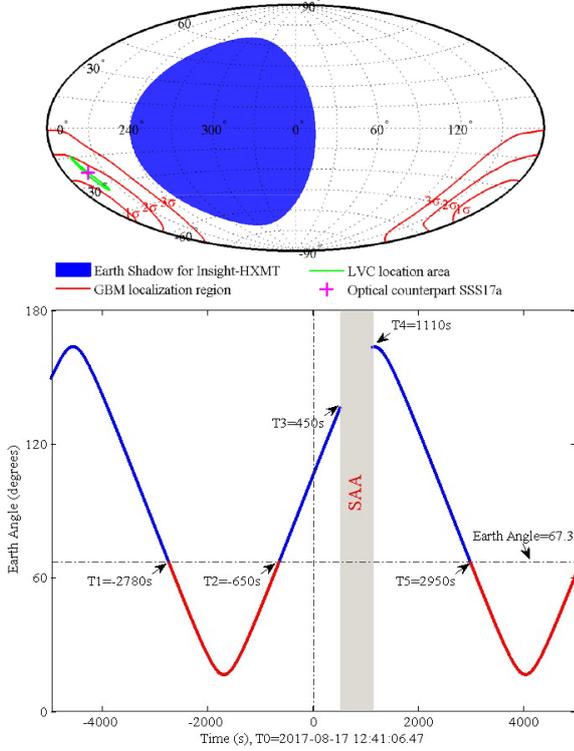

**Figure 4** (Top) The sky area monitored by HE for gamma-ray transients when GW170817 happened. The whole area of *Fermi*/GBM localization, LVC location and the optical counterpart SSS17a was not occulted by the Earth. (Bottom) Evolution of the Earth angle of the GW source within T0±5000 s. Earth angle < 67.3 means the source was occulted by the Earth. Gray area indicates the SAA passage when HE was turned off. Blue lines are time intervals when HE monitored the GW source (SSS17a).

Initial check of the light curve has revealed no counterpart of GRB170817A in the summed light curve of 18 CsI detectors, which was reported as one of the earliest GCNs regarding GW170817 [25]. Here the summed light curves with different temporal resolution (bin width) and time intervals are shown in Figure 5. The maximum time interval plotted in Figure 5 is T0-650 s and T0+450 s, during which *Insight*-HXMT provided the most valuable monitoring of any potential high energy emission from the GW source.

As shown in the top panel in Figure 5, there is no significant excess around the GW merger time or GRB trigger time [5]. We also did a Monte Carlo simulation for GRB170817A with the incident angle of the GW position and the spectrum measured by GBM [20], and found that some CsI detectors are expected to receive more counts than others (see Figure 6). Therefore, we tried different selection and combination of those CsI detectors in various energy ranges, but still did not find any excess. We also checked the light curve of NaI detectors and found no excess either. Non-detection of GRB170817A in both CsI and NaI is consistent with the expected counts (Figure 6) based on the GBM spectrum and the corrected detector response (see section 2.3). Thus we conclude that *Insight*-HXMT did not detect GRB170817A, suggesting that this burst is very weak in fluence and soft in spectrum, which is consistent with the GBM measurement [20].

---

[5] The time delay between *Insight*-HXMT and *Fermi*/GBM is ~3.56 ms for GRB170817A, which is negligible in this analysis.

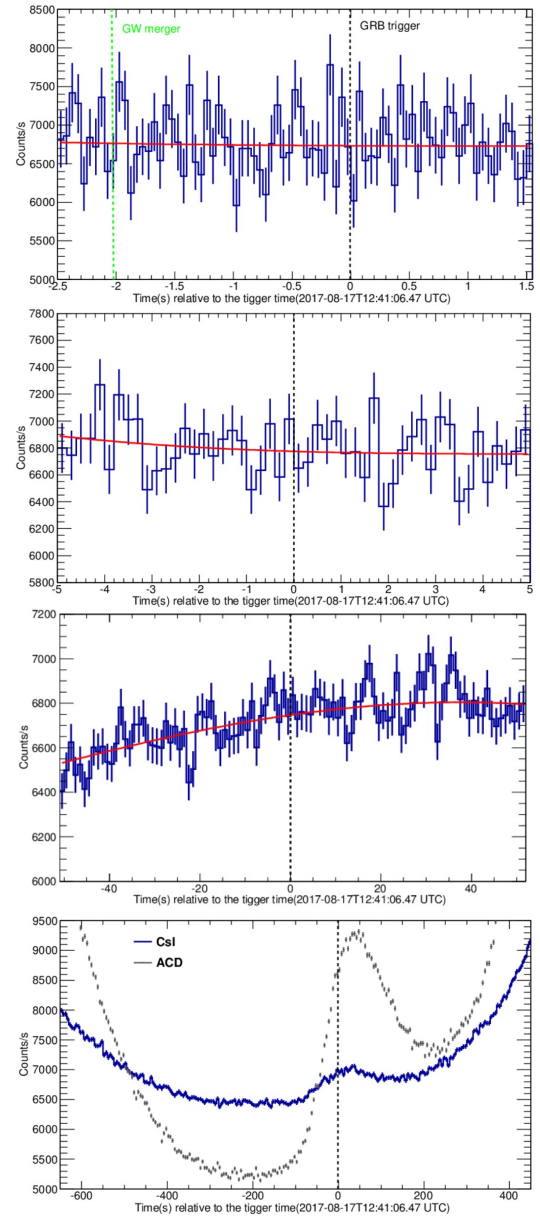

**Figure 5** Light curves of HE 18 CsI detectors around GW170817 and GRB 170817A. From top to bottom the time ranges and bin widths are: [T0-2.5 s, T0+1.5s] and 0.05 s; [T0-5 s, T0+5s] and 0.2 s; [T0-50 s, T0+50s] and 1 s; [T0-650 s, T0+450s] and 5 s, respectively. No significant excess above background variation are found. The bump from T0-100 s to T0+100 s is caused by charged particles in orbit, as indicated by the bump in the same time interval in the ACD light curve (gray)..

As shown in the bottom panel in Figure 5, a bump from T0-100 s to T0+100 s is evident. However, a similar bump is also seen in the same time interval in the plastic anticoincidence detectors (ACD), which are placed around the HE main detectors to reject the charged particles, in order to reduce the background of NaI/CsI [17]. Thus we conclude that this bump is caused by charged particles and has no relation with GW170817.

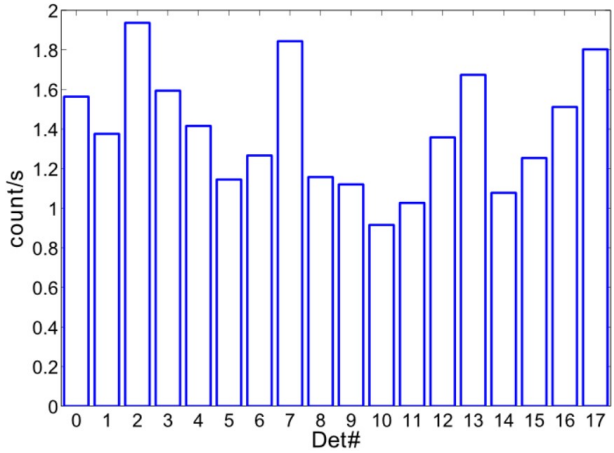

**Figure 6**  Simulated counts rate of 18 CsI detectors for GRB170817A using the GBM spectrum [20], GW incident angles and the corrected effective area (see section 2.3). In the 2 s duration of GRB170817A, the total count expected to be detected by HE is ~50, which is well within the 1-σ statistical fluctuation (~120) of the background in Figure 5.

Searches with time bins of 20 ms, 50 ms, 0.2 s and 1 s are also made to the raw light curve, and no significant excess with signal to noise ratio (SNR) greater than 5 is revealed; the small bumps in light curves are caused by the charged particles in orbit, which are usually seen in HE data.

### 2.2 ToO Observation

Immediately after receiving the alerts of GW170817 and GRB170817A, we scheduled a ToO observation to scan the localization region, which was originally provided by GBM and later by LIGO and Virgo collaboration (LVC) [23], aiming to find potential X-ray emission from the GW source. The ToO was actually implemented from T0+6.9 to T0+18.0 hour. During this ToO, the nominal FoVs of ME and LE did not cover the SSS17a area because it was too close to the Sun and beyond the sun avoidance angle (~70 deg) of the telescope. Carefully checking the observation history reveals that some LE detectors with very large FoV (~50 deg) [17] temporally covered the SSS17a area during the slew; however, those LE detectors were saturated due to the bright Earth.

### 2.3 Crab Calibration Observation

Since the response of the HE detector for GRB detection is very complicated and ground-calibration was not achieved, Monte Carlo simulation should be validated or corrected based on the in-flight calibration. Both the gamma-ray flux and the spectrum of the Crab pulsar are stable, which could be used to calibrate the effective area of HE CsI detector. We thus made a dedicated ToO observation with a total exposure time of 168 ks, for which the Crab was set in the same incident direction as GW170817.

The calibration using the Crab pulsar consists of the following steps: First, the arrival time of every event registered by CsI was converted to Solar System Barycenter (SSB) and the phase was obtained with the ephemeris of the Crab pulsar. Then the pulse profile was folded as shown in Figure 7. At the end, the spectrum accumulated from phase 0.6 to 0.8 was taken as the background for the pulsed component. Considering the exposure time, the mean pulsed flux is about 1.24±0.25 cnt/s.

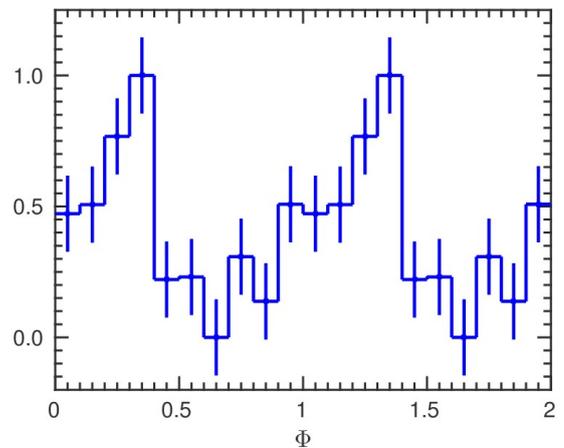

**Figure 7**  Crab pulse profile accumulated in the calibration observation, normalized with the maximum 1 and minimum 0. Phase 0 represents the position of the X-ray main peak observed by the HE NaI detector.

On the other hand, we calculated the expected mean count rate based on the spectrum of pulsed component of the Crab pulsar and the simulated detector response for the incident geometry of GW170817. The spectrum of the pulsed component is from [26]. This calculation gives a mean rate of 2.58 cnt/s, thus the ratio between observation and simulation is 0.48±0.1 (1 σ). The same analysis was applied to the observation of Crab in a different incident angle (theta=90.4 deg, phi=302.2 deg); the observed mean counts rate is 0.83±0.14 cnt/s while the simulated rate is 1.95, resulting in the ratio of 0.43±0.07 (1 σ), which is very close to the ratio for GW source. Therefore, we corrected the simulated effective area with the factor of 0.48 in the following analysis for the GW source.

## 3   Constraints on the high energy radiation from GW170817

According to short GRB studies in decades, the most likely high energy EM counterpart of BNS merger is the short GRB and its afterglow [14], which has been confirmed by the observation of GW170817 [23]. In addition, short GRBs could come with other radiation components, such as precursor before the trigger [27] or extended emission after the main pulse of GRB [28]. Although HE did not find any significant excess from GW170817 from T0-650 s to T0+450 s, *Insight*-HXMT/HE can pose stringent constraints (upper limits) on the high energy radiation from GW170817, including GRB170817A and other possible emission episodes.

We note that, among many operating gamma-ray telescopes, there are only four that monitored GW170817 during the trigger time (T0): *Fermi*/GBM, Konus-*Wind*, INTEGRAL/SPI-ACS and *Insight*-HXMT/HE [23]. Out of those four telescopes, HE has the unique advantage of large effective area (~1000 cm$^2$) as well as microsecond time resolution in the high energy band (0.2- 5 MeV).

To calculate the upper limit, we use the corrected detector response (see section 2.3) as well as different spectral models (see Table 1) and timescales for the potential emission components. Three typical spectral models for short GRB are used: Band function [29], Comptonized model (i.e. Power-Law with exponential cutoff) and simple Power-Law model. For each model, three sets of parameters are taken to represent the soft, medium and hard spectrum hardness. Spectral parameters are summarized in Table 1.

**Table 1** Spectral models are used for the upper limits calculation.

| Spectral model | alpha | beta | $E_{peak}$ (keV) |
|---|---|---|---|
| Band_1 | -1.9 | -3.7 | 70 |
| Band_2 | -1.0 | -2.3 | 230 |
| Band_3 | 0.0 | -1.5 | 1000 |
| Comp_1 | -1.9 | - | 70 |
| Comp_2 | -1.0 | - | 230 |
| Comp_3 | 0.0 | - | 1000 |
| PL_1 | -2.0 | - | - |
| PL_2 | -1.5 | - | - |
| PL_3 | -1.0 | - | - |

Since GRB170817A showed a strong spectral evolution [20], we computed the upper limits for two time intervals for the main burst: [-0.3 s, 0] and [0, 0.25 s]; all times are relative to the GBM trigger time (T0). For other potential emission episodes, timescales used to calculate upper limits are 0.1 s and 1 s for an assumed precursor before trigger, and 1 s and 10 s for the assumed extended emission after trigger. Detailed results of upper limits are shown in Figure 8 and summarized in Table 2. These tight upper limits down to ~$10^{-7}$ erg/cm$^2$/s in 0.2-5 MeV on the high energy radiation components including the prompt short GRB170817A and other possible precursor or extended emission will help us to better understand the properties of EM radiation from this double neutron star coalescence, as well as the physics of the relativistic jets of short GRBs.

## 4 Summary and Conclusions

GW170817 is the first BNS merger discovered by LIGO and Virgo. Fortunately its electromagnetic counterpart was successfully detected and identified in gamma-ray, optical, X-ray and radio band. *Insight*-HXMT/HE is one of the only four gamma-ray telescopes that successfully monitored GW170817 during the trigger time, and HE possesses the largest effective area (~1000 cm$^2$) in 0.2-5 MeV and highest time resolution (microsecond). Although HE did not detect GRB1708717A, it helped to confirm the unexpected weak and soft nature of this short GRB. Meanwhile, these observations made by HE allowed us to derive one of the most stringent constraints on GRB170817A and other possible emission components related to the GW source, which is very important to characterize the whole picture of this historical binary neutron star merger and its high energy electromagnetic counterpart.

*This work made use of the data from the Insight-HXMT mission, a project funded by China National Space Administration (CNSA) and the Chinese Academy of Sciences (CAS). The Insight-HXMT team gratefully acknowledges the support from the National Program on Key Research and Development Project (Grant No. 2016YFA0400800) from the Minister of Science and Technology of China (MOST) and the Strategic Priority Research Program of the Chinese Academy of Sciences (Grant No. XDB23040400). S.L. Xiong acknowledges the support from the Hundred Talent Program of Chinese Academy of Sciences. M. Y. Ge acknowledges the National Natural Science Foundation of China (NSFC), Grant No. 11233001 and 11503027. J. Zhang acknowledges the NSFC, Grant No. 11403026.*

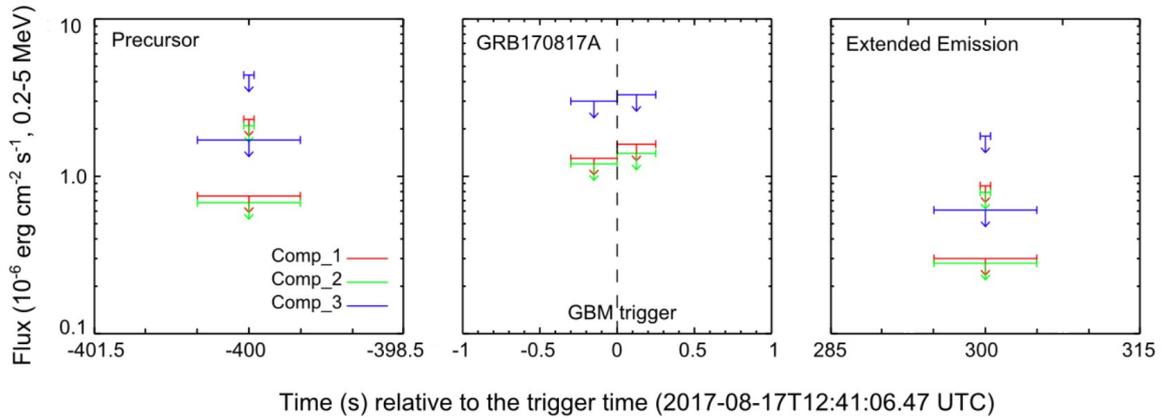

**Figure 8**  *Insight*-HXMT upper limits (3 σ) for GRB170817A, precursor and extended emission with various time-scales (width of the upper limit bar). Only results for three Comptonized models (see Table 1) are shown. See Table 2 for comprehensive upper limit results for all spectral models.

**Table 2**  *Insight*-HXMT upper limits (3 σ) for different time-scales and spectral models (see Table 1) for GRB170817A and other possible precursor and extended emission. T_start and T_end are the start and end time (relative to T0) of the assumed emission component to calculate the upper limits.

| Upper limits (erg/cm$^2$/s) | T_start (s) | T_end (s) | Band_1 | Band_2 | Band_3 | Comp_1 | Comp_2 | Comp_3 | PL_1 | PL_2 | PL_3 |
|---|---|---|---|---|---|---|---|---|---|---|---|
| GRB170817A | -0.30 | 0.00 | 1.4e-06 | 1.9e-06 | 3.4e-06 | 1.3e-06 | 1.2e-06 | 3.0e-06 | 2.2e-06 | 2.8e-06 | 2.1e-06 |
| GRB170817A | 0.00 | 0.25 | 1.6e-06 | 2.2e-06 | 3.6e-06 | 1.6e-06 | 1.4e-06 | 3.3e-06 | 2.6e-06 | 3.1e-06 | 2.3e-06 |
| Precursor | -400.05 | -399.95 | 2.3e-06 | 3.1e-06 | 4.8e-06 | 2.3e-06 | 2.1e-06 | 4.4e-06 | 3.6e-06 | 4.1e-06 | 2.9e-06 |
| Precursor | -400.5 | -399.5 | 7.7e-07 | 1.1e-06 | 1.8e-06 | 7.5e-07 | 6.8e-07 | 1.7e-06 | 1.3e-06 | 1.5e-06 | 1.1e-06 |
| Extended | 299.5 | 300.5 | 8.9e-07 | 1.2e-06 | 2.0e-06 | 8.7e-07 | 7.9e-07 | 1.8e-06 | 1.4e-06 | 1.8e-06 | 1.2e-06 |
| Extended | 295 | 305 | 3.1e-07 | 4.3e-07 | 6.4e-07 | 3.0e-07 | 2.8e-07 | 6.1e-07 | 5.2e-07 | 6.1e-07 | 4.1e-07 |